\documentclass[10pt]{article}

\usepackage[a4paper,margin=1in]{geometry}
\usepackage{amsmath,amssymb,amsfonts}
\usepackage{booktabs}
\usepackage{array}
\usepackage{graphicx}
\usepackage{multirow}
\usepackage{caption}
\usepackage{subcaption}
\usepackage{bm}
\usepackage{physics}
\usepackage{hyperref}
\usepackage{tabularx}
\usepackage{ragged2e}
\usepackage{orcidlink}
\usepackage{xcolor}

\hypersetup{
    colorlinks=true,
    linkcolor=blue,
    citecolor=blue,
    urlcolor=blue
}

\newcolumntype{Y}{>{\RaggedRight\arraybackslash}X}

\title{From Scalar $H_0$ to $E(z)$: A Reformulation of the Hubble Tension}

\author{
Seokcheon Lee\,\orcidlink{0000-0003-0861-1300}\\
Department of Physics, Institute of Basic Science, Sungkyunkwan University,\\
Suwon 16419, Republic of Korea\\
\texttt{skylee@skku.edu}
}

\date{}

\begin{document}

\maketitle

\begin{abstract}
The Hubble tension is usually expressed as a discrepancy between the low $H_0$ inferred from Planck CMB data within base $\Lambda$CDM and the higher value obtained from late-time distance-ladder measurements. This scalar comparison compresses distinct inference problems into one derived parameter: Planck CMB, DESI DR2 BAO, and Pantheon$+$SH0ES constrain physical densities and acoustic scales, ruler-normalized distances, and calibrated luminosity-distance relations, respectively. We reformulate the comparison in terms of the dimensionless expansion history $E(z)=H(z)/H_0$. This does not remove the absolute-scale discrepancy, but separates the normalization encoded in $H_0$ from the redshift-dependent shape of the expansion history. Within a common flat-$\Lambda$CDM framework, each probe posterior is mapped onto posterior-implied $E(z)$ histories. Since the reconstructed values $E(z_k)$ are strongly correlated across redshift, we quantify the global mismatch with a covariance-subspace history displacement $S_{\rm hist}$, alongside pointwise redshift differences. The histories are not identical, but the discrepancies are moderate: the pointwise significance is typically $1$--$2\sigma$, while $S_{\rm hist}\simeq1.65$ for DESI DR2 and $S_{\rm hist}\simeq2.55$ for Pantheon$+$SH0ES relative to Planck. With two retained covariance eigenmodes, these correspond to two-sided one-dimensional Gaussian equivalents of approximately $1.1\sigma$ and $2.1\sigma$, both below the conventional $\simeq4.9\sigma$ Planck--SH0ES scalar-$H_0$ discrepancy.
\end{abstract}

%==========================================================
\section{Introduction}
\label{sec:intro}
%==========================================================

A central open issue in precision cosmology is the discrepancy between the Planck-inferred value of $H_0$ in base $\Lambda$CDM and the higher values obtained from late-time distance measurements~\cite{Planck:2018vyg,Verde:2019ivm,DiValentino:2021izs,Perivolaropoulos:2021jda,Riess:2021jrx,Abdalla:2022yfr}. Independent late-time routes, including Cepheid-calibrated SNe Ia, TRGB calibrations, time-delay lenses, and megamasers, have sharpened the empirical status of the problem~\cite{Riess:2021jrx,H0LiCOW:2019pvv,Freedman:2019jwv,Pesce:2020xfe}. Although a scalar comparison in $H_0$ is useful as a compact summary, it compresses different inference problems into a single derived parameter. CMB, BAO, and distance-ladder analyses do not directly constrain the same primitive observable. Their quoted values of $H_0$ are obtained via different inference chains involving acoustic scales, ruler-normalized distances, and calibrated luminosity distances~\cite{Planck:2018vyg,Riess:2021jrx,Knox:2019rjx,Efstathiou:2021ocp,Brieden:2022heh,DESI:2025zgx}.

The distinction is evident in the quantities most directly constrained by each probe. Planck CMB data constrain physical densities and acoustic angular scales, including $\omega_b$, $\omega_c$, and $\theta_\ast$. Here $\theta_\ast$ encodes the ratio of the sound horizon to the angular-diameter distance to last scattering~\cite{Planck:2018vyg,Hu:1995en,Hu:2001bc}. DESI DR2 BAO constrains standard-ruler-normalized distances, including $D_M(z)/r_d$ and $D_H(z)/r_d$ in anisotropic bins, and $D_V(z)/r_d$ in isotropic bins~\cite{DESI:2025zgx,SDSS:2005xqv,eBOSS:2020hur}. Pantheon$+$ constrains the Type Ia supernova luminosity-distance relation through standardized distance moduli over a broad redshift range~\cite{SDSS:2009pcx,SDSS:2014iwm,Pan-STARRS1:2017jku,Scolnic:2021amr,Brout:2022vxf}. When combined with SH0ES, the absolute scale is fixed by Cepheid-calibrated supernova hosts, yielding a calibrated distance ladder and an inferred local value of $H_0$~\cite{Riess:2021jrx,Riess:2016jrr,Riess:2019cxk}. These probe-level quantities are not identical and do not enter the inference problem in the same way.

Therefore, the quoted scalar $H_0$ should be treated as a derived late-time summary, not as a common primary observable~\cite{Verde:2019ivm,Knox:2019rjx,Efstathiou:2021ocp,Verde:2023lmm,Poulin:2024ken}. For Planck, $H_0$ is obtained by projecting CMB acoustic information through a background cosmological model~\cite{Planck:2018vyg,Knox:2019rjx}. For BAO, it is inferred only after standard-ruler-normalized distances are interpreted with assumptions about $r_d$ and the background expansion~\cite{Efstathiou:2021ocp,DESI:2025zgx,Poulin:2024ken}. For the distance ladder, it is tied to the absolute calibration of the supernova magnitude-redshift relation~\cite{Riess:2021jrx,Scolnic:2021amr,Brout:2022vxf}. Thus, the same symbol $H_0$ denotes quantities obtained through different inference chains.

Differences among inferred late-time parameters are therefore not automatically differences among the directly constrained quantities. CMB, BAO, and distance-ladder data constrain different primitive combinations, and the map from those combinations to parameters such as $\Omega_{m0}$ and $H_0$ is model dependent~\cite{Knox:2019rjx,Efstathiou:2021ocp,Poulin:2024ken,Pedrotti:2024kpn,Maus:2024dzi}. A related information-geometric formulation emphasized that changes in the model manifold can rotate parameter-degeneracy directions and modify the apparent sharpness of the inferred $H_0$ constraint~\cite{Lee:2026ets}. In extended models, the map also involves late-time degrees of freedom such as the dark-energy equation of state. A disagreement in derived parameters can reflect either a mismatch in the reconstructed background history or a difference in projection geometry. This distinction motivates a comparison in a common dimensionless history space.

This perspective is complementary to earlier discussions of CMB-inferred
$H_0$, sound-horizon calibration, and model-dependent projection
~\cite{Efstathiou:2021ocp,Brieden:2022heh}. Those works clarify why the quoted
scalar $H_0$ is not a primitive observable. Here we ask a different question:
after the Planck, DESI DR2, and Pantheon$+$SH0ES posteriors are mapped into a
common flat-$\Lambda$CDM background, do they imply a comparably large mismatch
in the normalized expansion history $E(z)$?

A natural object for this comparison is $E(z)\equiv H(z)/H_0$. Instead of asking only whether different probes return the same scalar value of $H_0$, one may ask whether they imply mutually consistent normalized expansion histories. The use of $E(z)$ does not divide away the absolute-scale discrepancy. It separates the absolute normalization encoded in $H_0$ from the redshift-dependent shape of the late-time expansion history. In this sense, the comparison in $E(z)$ is a scale--shape decomposition of the background expansion. Similar dimensionless diagnostics appear in consistency tests based on $E(z)$, the Om diagnostic, ruler-normalized BAO distances such as $D_M(z)/r_d$ and $D_H(z)/r_d$, and the dimensionless luminosity-distance combination $H_0D_L(z)/c$~\cite{DESI:2025zgx,SDSS:2005xqv,Scolnic:2021amr,Brout:2022vxf,Hogg:1999ad,Clarkson:2007pz,Sahni:2008xx,Shafieloo:2012ht,eBOSS:2020yzd}. In this paper, we focus on $E(z)$ because it directly represents the normalized background expansion history.

Several recent works have moved beyond a one-dimensional comparison of quoted $H_0$ values. Some analyses classify or combine determinations of $H_0$ that are independent of the sound horizon, while others study inverse-distance-ladder or calibration-based routes to the expansion scale~\cite{Efstathiou:2021ocp,Poulin:2024ken,Bernal:2016gxb,Aylor:2018drw,Perivolaropoulos:2024yxv,Pantos:2026cxv}. Perivolaropoulos identified the main scalar discrepancy as one between distance-ladder measurements and other determinations~\cite{Perivolaropoulos:2024yxv}. Pantos and Perivolaropoulos reached a similar conclusion using a larger set of sound-horizon-free measurements~\cite{Pantos:2026cxv}. These works classify scalar $H_0$ determinations. They do not compare posterior-implied expansion histories in the covariance subspace of $E(z)$.

Other studies reconstruct or test the late-time expansion history more directly. Jiang \textit{et al.}\ reconstructed $E(z)$ nonparametrically using DESI BAO and Type Ia supernova data, and discussed its implications for the Hubble tension~\cite{Jiang:2024xnu}. Flexible and nonparametric approaches have also been used to reconstruct dark-energy dynamics and late-time expansion histories beyond fixed parametric forms~\cite{Huang:2024erq,Ormondroyd:2025exu}. Bansal and Huterer studied the expansion-history preference of DESI DR2 and external data~\cite{Bansal:2025ipo}. DESI-era analyses have also tested interacting and nonlinear dark-sector extensions against late-time background constraints~\cite{Figueruelo:2026eis}. Null-test analyses use $E(z)$, its derivatives, and related diagnostics to test $\Lambda$CDM and FLRW consistency without committing to a specific dark-energy parametrization~\cite{Clarkson:2007pz,Sahni:2008xx,Shafieloo:2012ht,Dinda:2025svh}.

The present analysis has a different target. It does not seek a nonparametric best-fit expansion history, nor does it test a specific deformation of $\Lambda$CDM. Instead, it performs a controlled common-model projection test. BAO and SNe data constrain distances, calibrated distance moduli, and ruler-normalized quantities; they do not directly measure $E(z)$ at independent redshifts~\cite{DESI:2025zgx,SDSS:2005xqv,Scolnic:2021amr,Brout:2022vxf,Hogg:1999ad}. The question is therefore limited: when the Planck, DESI DR2, and Pantheon$+$SH0ES posteriors are interpreted within the same flat base $\Lambda$CDM background, what normalized expansion histories do they imply, and how large is the corresponding covariance-aware history-space mismatch?

This approach is related to recent work emphasizing that the Hubble tension is not a purely one-dimensional problem in $H_0$. Poulin \textit{et al.}\ formulated the related ``cosmic calibration tension'' as a discrepancy that extends beyond $H_0$ and involves the consistency of BAO, SNe Ia, and CMB calibration routes~\cite{Poulin:2024ken}.  Pedrotti \textit{et al.}\ emphasized the multidimensional character of the tension and the roles of $\Omega_m$ and the physical cold-dark-matter density $\omega_c$~\cite{Pedrotti:2024kpn}. Similar caution appears in analyses of early- and late-time calibration, sound-horizon dependence, and multi-parameter consistency tests~\cite{Knox:2019rjx,Efstathiou:2021ocp,Verde:2023lmm,Bernal:2016gxb,Aylor:2018drw,Schoneberg:2021qvd}. These works motivate caution in treating a scalar $H_0$ comparison as a complete diagnostic of cosmological consistency.

A related methodological issue is parameter compression. Cosmological likelihoods are often summarized by compressed parameters before being projected onto derived cosmological quantities. This practice is useful, but it can hide the geometry of the original likelihood and the correlations among derived parameters~\cite{Tegmark:1997rp,Heavens:2009nx,Alsing:2018eau,Charnock:2018ogm}. In the BAO context, compressed distance measurements and full-shape or full-modeling analyses do not always encode the same information~\cite{DESI:2025zgx,Maus:2024dzi,eBOSS:2020yzd,DESI:2024mwx}. Maus \textit{et al.}\ compared parameter-compression and full-modeling approaches for DESI and showed how the chosen compression can affect the interpretation of cosmological constraints~\cite{Maus:2024dzi}. The present paper does not perform a separate full-likelihood versus compressed-parameter test. Instead, it asks a narrower question: whether the same probe posteriors show a smaller mismatch when compared in the function space of $E(z)$ than when summarized by the scalar $H_0$.

We find that the reconstructed histories are not identical, but their discrepancies are milder than the conventional scalar-$H_0$ comparison. Once the redshift covariance of the reconstructed histories is included, the history-space mismatch remains at a moderate level. This result does not remove the absolute-scale tension. It shows that the scalar-$H_0$ discrepancy and the normalized-history discrepancy are different statistical questions, consistent with recent discussions of multidimensional and calibration-based formulations of the Hubble tension~\cite{Knox:2019rjx,Efstathiou:2021ocp,Poulin:2024ken,Pedrotti:2024kpn}.

The paper is organized as follows. Section~\ref{sec:2} distinguishes the directly constrained probe-level quantities from the displayed scalar summaries commonly quoted in the literature. In Section~\ref{sec:3}, we present the reconstruction of $E(z)$ and the covariance-aware comparison of the resulting histories. Section~\ref{sec:4} discusses the implications for interpreting the Hubble tension. Section~\ref{sec:5} summarizes the conclusions and comments on extensions beyond flat $\Lambda$CDM.

%==========================================================
\section{Probe-Level Observables and the Status of Quoted $H_0$}
\label{sec:2}
%==========================================================

The comparison developed in this paper starts from a simple distinction. Planck CMB, DESI DR2 BAO, and Pantheon$+$SH0ES do not directly constrain the same primitive observable. Each probe constrains a different set of probe-level quantities. The quoted scalar $H_0$ is obtained only after those quantities are interpreted within a background cosmological model.

Therefore, we do not treat the quoted values of $H_0$ as direct measurements of the same object. Instead, we separate the quantities most directly constrained by each probe from the late-time scalar summaries derived from them. This separation motivates the use of posterior-implied dimensionless histories, such as $E(z)$, as a common comparison space.

The use of $E(z)$ should be understood in this restricted sense. It is not a claim that the absolute scale $H_0$ is irrelevant. It is also not a model-independent reconstruction of the expansion history from the data alone. Rather, $E(z)$ isolates the normalized redshift-dependent shape of the background expansion after the absolute scale has been factored out. Comparing posterior-implied $E(z)$ histories therefore asks whether the probes disagree in the shape of the late-time expansion history, rather than only in the absolute normalization summarized by $H_0$.

This use of $E(z)$ is complementary to earlier discussions of CMB-inferred
$H_0$, sound-horizon calibration, and model-dependent projection
~\cite{Efstathiou:2021ocp,Brieden:2022heh}. Those works focus on how the
quoted scalar $H_0$ depends on calibration, the sound horizon, and the assumed
background model. The present comparison instead asks how the same
probe-level information appears after being mapped into the normalized
background history $E(z)$ within a common flat-$\Lambda$CDM framework.

%----------------------------------------------------------
\subsection{Planck CMB}
%----------------------------------------------------------

For Planck CMB, the late-time scalar $H_0$ is not directly observed. The CMB anisotropy spectra constrain physical densities and acoustic angular scales. In base $\Lambda$CDM, the relevant quantities include
\begin{equation}
\omega_b\equiv \Omega_{b0}h^2,\qquad \omega_c\equiv \Omega_{c0}h^2,\qquad \theta_\ast .  \label{eq:planck_primary_params}
\end{equation}
The acoustic angular scale is
\begin{equation}
\theta_\ast = \frac{r_s(z_\ast)}{D_M(z_\ast)} , \label{eq:theta_ast}
\end{equation}
where $r_s(z_\ast)$ is the comoving sound horizon at photon decoupling and $D_M(z_\ast)$ is the transverse comoving distance to last scattering~\cite{Planck:2018vyg,Hu:1995en,Hu:2001bc}. The numerator is controlled mainly by early-universe physics. The denominator depends on the integrated expansion history between last scattering and today.

Thus, in Planck CMB analyses, $H_0$ is inferred through a model-dependent projection. Within flat base $\Lambda$CDM, the CMB posterior implies distributions for $\Omega_{m0}$ and $H_0$, and hence for
\begin{equation}
E(z)= \left[ \Omega_{m0}(1+z)^3+1-\Omega_{m0} \right]^{1/2}. \label{Ez}
\end{equation}
The resulting $E(z)$ history is posterior-implied within the model. It is not a direct reconstruction from CMB data alone.

%----------------------------------------------------------
\subsection{DESI DR2 BAO}
%----------------------------------------------------------

DESI DR2 BAO also does not directly measure a standalone Hubble constant. BAO measurements constrain distances relative to the sound horizon at the drag epoch. The relevant observables include
\begin{equation}
\frac{D_M(z)}{r_d}, \qquad \frac{D_H(z)}{r_d}, \label{eq:bao_aniso}
\end{equation}
in anisotropic BAO measurements, and
\begin{equation}
\frac{D_V(z)}{r_d}
\label{eq:bao_iso}
\end{equation}
in isotropic measurements~\cite{DESI:2025zgx,SDSS:2005xqv,eBOSS:2020yzd}. Here
\begin{equation}
D_H(z)=\frac{c}{H(z)}
\label{eq:DH_def}
\end{equation}
is the Hubble distance, and $r_d$ is the sound horizon at the drag epoch.

For a flat FLRW background,
\begin{equation}
D_H(z)=\frac{c}{H_0E(z)}, \qquad D_M(z)=\frac{c}{H_0}\int_0^z \frac{dz'}{E(z')}. \label{eq:DM_DH_flat}
\end{equation}
Thereforer, the BAO likelihood constrains combinations of the expansion history, the distance scale, and the ruler scale. A quoted value of $H_0$ derived from BAO is not a primitive BAO observable. It is inferred only after specifying the background model and the treatment of $r_d$.

This distinction is important because not all BAO combinations carry the same dependence on $r_d$. Quantities such as $D_M/r_d$, $D_H/r_d$, and $D_V/r_d$ are ruler-normalized, whereas ratios such as $D_M/D_H$ are independent of $r_d$ at the level of the BAO observable. This separation is useful when assessing how BAO information is projected into late-time parameters~\cite{Brieden:2022heh,Lee:2025ysg,Lee:2025jrk}. In the present paper, any $E(z)$ history assigned to the DESI DR2 posterior is obtained only after choosing the flat $\Lambda$CDM background. Thus, it is a posterior-implied history, not a model-independent BAO reconstruction of $H(z)/H_0$ at independent redshifts.

%----------------------------------------------------------
\subsection{Pantheon$+$SH0ES}
%----------------------------------------------------------

Pantheon$+$SH0ES constrains the luminosity-distance relation with an absolute calibration. Pantheon$+$ supernovae determine the relative distance-redshift relation over a broad redshift range~\cite{Scolnic:2021amr,Brout:2022vxf}. By itself, the supernova Hubble diagram does not fix $H_0$ independently of the absolute magnitude calibration. The apparent magnitude depends on the combination of the supernova absolute magnitude and the distance scale.

Using the flat-FLRW relation $D_L(z)=(1+z)D_M(z)$ and Eq.~\eqref{eq:DM_DH_flat}, one obtains
\begin{equation}
D_L(z)= \frac{c}{H_0}(1+z)\int_0^z \frac{dz'}{E(z')}. \label{DL}
\end{equation}
Thus, Type Ia supernovae constrain an integral of the expansion history rather than $H(z)$ directly. SH0ES supplies the Cepheid-based calibration of the supernova absolute magnitude. This calibration fixes the absolute distance scale and yields a local value of $H_0$~\cite{Riess:2021jrx}. The resulting $H_0$ is tied to the calibrated distance ladder. It is not the same type of derived quantity as the CMB-inferred $H_0$ or a BAO-derived $H_0$.

Pantheon$+$SH0ES therefore supplies two ingredients: a calibrated absolute scale and a relative distance-redshift relation. When mapped into $E(z)$ within flat $\Lambda$CDM, the comparison tests the normalized shape implied by this distance-redshift relation and its calibration. It does not provide an independent direct measurement of $H(z)/H_0$ at each redshift.

%----------------------------------------------------------
\subsection{Implication}
%----------------------------------------------------------

The three probes enter the inference problem through different observables. Planck CMB constrains acoustic scales and physical densities. DESI DR2 BAO constrains standard-ruler-normalized distances. Pantheon$+$SH0ES constrains a calibrated luminosity-distance relation. In all three cases, the quoted scalar $H_0$ is a derived late-time summary.

This does not make the scalar $H_0$ comparison irrelevant. It means that the comparison is not a direct comparison of a common primitive observable. A background-history comparison asks a different question: whether the same probe posteriors imply mutually consistent dimensionless expansion histories after they are mapped into a common model space.

The comparison in $E(z)$ should therefore be viewed as a scale--shape decomposition, not as a replacement for the scalar-$H_0$ comparison. The scalar $H_0$ comparison diagnoses the absolute normalization of the late-time distance scale. The $E(z)$ comparison diagnoses the normalized redshift-dependent shape of the expansion history. If the scalar comparison shows a large discrepancy while the $E(z)$ comparison does not, the tension is localized mainly in the absolute-scale projection rather than in the late-time dynamical shape. For this reason, the following sections use reconstructed $E(z)$ histories as the primary normalized-history comparison space, while retaining the scalar $H_0$ discrepancy as the reference absolute-scale comparison.

Table~\ref{tab:probe_observables} summarizes the probe-level quantities and the role of the quoted scalar $H_0$ in each case.

\begin{table*}[t]
\centering
\caption{
Probe-level quantities entering the three data sets considered in this work.
The quoted scalar $H_0$ is not a common primitive observable; it is a
probe-dependent derived summary obtained after model interpretation or
calibration.
}
\label{tab:probe_observables}
\renewcommand{\arraystretch}{1.15}
\setlength{\tabcolsep}{6pt}
\begin{tabularx}{\textwidth}{>{\RaggedRight\arraybackslash}p{2.7cm} Y Y}
\toprule
Probe
& Main probe-level information
& Status of quoted $H_0$ \\
\midrule

Planck CMB
& Physical densities and acoustic angular scale:
$\omega_b$, $\omega_c$, $\theta_\ast=r_s(z_\ast)/D_M(z_\ast)$
& Inferred by projecting CMB acoustic information through a background
cosmological model \\

DESI DR2 BAO
& Standard-ruler-normalized distances:
$D_M(z)/r_d$, $D_H(z)/r_d$, and $D_V(z)/r_d$ where applicable
& Inferred only after specifying the background model and the treatment of
$r_d$ \\

Pantheon$+$SH0ES
& Relative SN Ia distance-redshift relation plus Cepheid-calibrated absolute
magnitude
& Follows from the absolute normalization of the calibrated distance ladder \\

\bottomrule
\end{tabularx}
\end{table*}

%==========================================================
\section{Reconstructed $E(z)$ Histories and Covariance-Aware Comparison}
\label{sec:3}
%==========================================================

We now compare the probes in a common dimensionless history space. All reconstructions are performed within flat base $\Lambda$CDM. This fixes the background model and places the three probe posteriors on the same comparison basis.

This is not a model-independent reconstruction of $E(z)$. Such reconstructions
have been studied elsewhere, for example using DESI BAO and Type Ia supernova
data~\cite{Jiang:2024xnu}. Here the purpose is more limited. Since the
conventional Planck--SH0ES Hubble tension is usually stated within base
$\Lambda$CDM, we ask whether the same flat-$\Lambda$CDM projection also
produces a comparably large mismatch in the normalized expansion history.
DESI DR2 BAO constrains combinations such as $D_M(z)/r_d$, $D_H(z)/r_d$, and
$D_V(z)/r_d$, while Pantheon$+$SH0ES constrains a calibrated luminosity-distance
relation. The mapping from these observables to $E(z)$ is performed only after
choosing the flat-$\Lambda$CDM background. The reconstructed histories are
therefore posterior-implied histories within a fixed model, not direct
measurements of $E(z)$ in independent redshift bins.

In flat $\Lambda$CDM, the dimensionless expansion history is given by Eq.~\eqref{Ez}. The reconstructed late-time history is therefore determined by the posterior distribution of $\Omega_{m0}$. For Planck, we use the published base-$\Lambda$CDM posterior chains. For DESI DR2 and Pantheon$+$SH0ES, we use the flat-$\Lambda$CDM Gaussian posterior approximations described in Appendix~\ref{app:numerical_method}. Each posterior sample is mapped to an $E(z)$ curve on a common redshift grid.

\begin{figure*}[!t]
\centering
\includegraphics[width=0.82\textwidth]{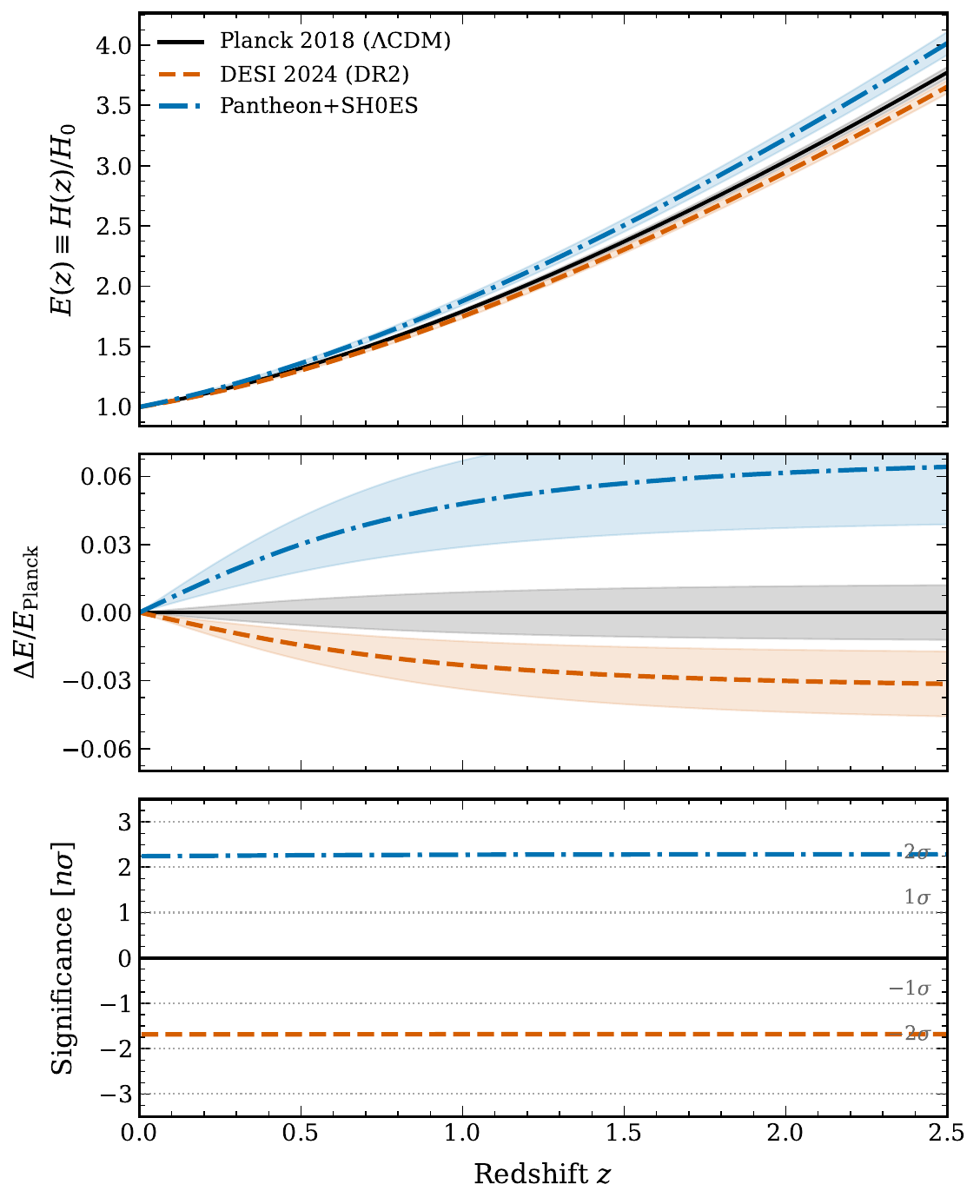}
\caption{Reconstructed dimensionless expansion histories in flat base $\Lambda$CDM. The top panel shows the central $E(z)$ histories for Planck 2018, DESI DR2, and Pantheon$+$SH0ES. The middle panel shows the fractional difference relative to the Planck-implied history, $\Delta E(z)/E_{\rm Planck}(z)$, with uncertainty bands. The bottom panel shows the pointwise significance $S(z)$ defined in Eq.~\eqref{eq:pointwise_significance}; the horizontal dotted lines mark the $\pm1\sigma$ and $\pm2\sigma$ levels.}
\label{fig:Ez_reconstruction}
\end{figure*}

Figure~\ref{fig:Ez_reconstruction} shows the comparison. The top panel shows the central reconstructed $E(z)$ histories. The middle panel shows the fractional difference relative to Planck,
\begin{equation}
\frac{\Delta E_i(z)}{E_{\rm Planck}(z)} = \frac{\bar E_i(z)-\bar E_{\rm Planck}(z)}{\bar E_{\rm Planck}(z)} , \label{eq:frac_difference}
\end{equation}
with uncertainty bands. The bottom panel shows the pointwise significance,
\begin{equation}
S_i(z)= \frac{\bar E_i(z)-\bar E_{\rm Planck}(z)}{\sqrt{\sigma_i^2(z)+\sigma_{\rm Planck}^2(z)}} , \label{eq:pointwise_significance}
\end{equation}
where $i$ denotes DESI DR2 or Pantheon$+$SH0ES. The quantity $\sigma(z)$ is estimated from the corresponding $68\%$ posterior interval.

The reconstructed histories are not identical. Relative to Planck, the DESI DR2 history lies below the Planck-implied curve over the plotted redshift range. The Pantheon$+$SH0ES history lies above it. Thus the probes show a visible offset even in the dimensionless history variable $E(z)$.

The pointwise mismatch is modest. The maximum fractional difference is about $3.1\%$ for DESI DR2 and about $6.4\%$ for Pantheon$+$SH0ES relative to Planck. The corresponding maximum pointwise significances are about $1.68\sigma$ and $2.29\sigma$. These numbers are local diagnostics on the chosen redshift grid. They should not be read as independent redshift-bin tensions. In flat $\Lambda$CDM, the reconstructed $E(z)$ curve is driven mainly by $\Omega_{m0}$, so the redshift points are strongly correlated.

A diagonal redshift-by-redshift statistic would ignore this correlation
\begin{equation}
\chi_{E,\mathrm{diag}}^2 = \sum_k \frac{ \left[ \bar E_i(z_k)-\bar E_{\rm Planck}(z_k)\right]^2}{
\sigma_i^2(z_k)+\sigma_{\rm Planck}^2(z_k)}. \label{eq:chi2_diag}
\end{equation}
This gives the apparent values
\begin{equation}
S_{\rm diag}\simeq 33.6 \quad \text{for DESI DR2 vs Planck}, \qquad
S_{\rm diag}\simeq 45.4 \quad \text{for Pantheon$+$SH0ES vs Planck}.
\label{eq:Sdiag_values}
\end{equation}
We do not interpret these numbers as physical significances. They are included only as a diagnostic warning. The diagonal sum treats the redshift grid points as independent measurements, although in flat $\Lambda$CDM they are coherent projections of the same low-dimensional posterior. The large values of $S_{\rm diag}$ therefore quantify over-counting by the diagonal approximation, not an inter-probe tension at the quoted level.

Thus, we use the posterior-induced redshift covariance of the reconstructed histories. For two independent probes, the covariance of the history difference is
\begin{equation}
C_{\Delta E}=C_{E,i}+C_{E,\rm Planck}. \label{eq:Cdelta_main}
\end{equation}
The covariance-aware statistic is
\begin{equation}
\chi_E^2 = \Delta{\bf E}^{\mathsf T} C_{\Delta E}^{+} \Delta{\bf E}, \qquad \Delta{\bf E} = \bar{\bf E}_i-\bar{\bf E}_{\rm Planck} \,,
\label{eq:chi2_cov}
\end{equation}
where $C_{\Delta E}^{+}$ denotes the inverse on the retained covariance subspace. The construction of $C_E$ from posterior samples and the eigenmode truncation procedure are described in Appendix~\ref{app:numerical_method}.

The covariance spectrum is nearly rank one. In the fiducial calculation we retain two eigenmodes, but we also track the leading-mode contribution to test the stability of the result against the retained-rank choice. We define the covariance-subspace history displacement as
\begin{equation}
S_{\rm hist} \equiv \sqrt{\chi_E^2}. \label{eq:Shist_def}
\end{equation}
The resulting values are
\begin{equation}
S_{\rm hist}\simeq 1.65 \quad \text{for DESI DR2 vs Planck}, \qquad
S_{\rm hist}\simeq 2.55 \quad \text{for Pantheon$+$SH0ES vs Planck}.
\label{eq:cov_significance_results}
\end{equation}

The quantity $S_{\rm hist}$ is a distance in the retained covariance subspace. It should not be identified with a one-dimensional Gaussian number of sigma. If the retained modes are treated as Gaussian degrees of freedom, then $S_{\rm hist}^2$ is compared with a $\chi^2$ distribution with $\nu=N_{\rm kept}$ degrees of freedom. The corresponding tail probability is
\begin{equation}
p_{\rm hist} = P\!\left(\chi^2_{\nu}\ge S_{\rm hist}^2\right), \qquad \nu=N_{\rm kept}. \label{eq:p_hist}
\end{equation}
When useful, this tail probability can be converted to a two-sided one-dimensional Gaussian equivalent $Z_{\rm 1D}$ through
\begin{equation}
p_{\rm hist} = 2\left[1-\Phi(Z_{\rm 1D})\right], \label{eq:Z1D_equiv}
\end{equation}
where $\Phi$ is the cumulative distribution function of a standard normal variable. The statistical meaning of this conversion, and its distinction from a one-dimensional Gaussian density, is summarized in Appendix~\ref{app:gaussian_equivalent}.

In the fiducial two-mode covariance-subspace calculation, $N_{\rm kept}=2$. Thus, $S_{\rm hist}=1.65$ corresponds to $\chi_E^2\simeq 2.72$, $p_{\rm hist}\simeq 0.26$, or a two-sided one-dimensional Gaussian equivalent of about $1.1\sigma$. Similarly, $S_{\rm hist}=2.55$ corresponds to $\chi_E^2\simeq 6.50$, $p_{\rm hist}\simeq 0.039$, or a two-sided one-dimensional Gaussian equivalent of about $2.1\sigma$. These Gaussian-equivalent values are smaller than the conventional $\sim4.9\sigma$ scalar Planck--SH0ES discrepancy~\cite{Planck:2018vyg,Verde:2019ivm,Riess:2021jrx}. They are also much smaller than the diagonal over-counting diagnostic in Eq.~\eqref{eq:Sdiag_values} and are consistent with the pointwise behavior shown in Fig.~\ref{fig:Ez_reconstruction}. As shown below, the qualitative conclusion is unchanged when the comparison is restricted to the leading covariance mode.

Table~\ref{tab:cov_modes} summarizes the covariance-eigenmode calculation.
The covariance spectrum is strongly hierarchical in both comparisons, as shown
by the small ratios $\lambda_2/\lambda_1$. Thus, $N_{\rm kept}=2$ should not be
interpreted as evidence for two independent redshift-bin measurements. It is a
regularized covariance-subspace choice. The leading-mode fraction shows how
much of $S_{\rm hist}^2$ is carried by the dominant covariance mode.

\begin{table}[t]
\centering
\caption{
Covariance-eigenmode summary for the reconstructed $E(z)$ mismatch relative to
Planck. The eigenvalues are ordered from largest to smallest. In the fiducial
calculation two modes are retained, so $\lambda_2/\lambda_1$ measures the
hierarchy of the retained covariance spectrum. The quantities
$S_{\rm hist}^{(2)}$ and $S_{\rm hist}^{(1)}$ denote the covariance-subspace
displacement obtained with two retained modes and with the leading mode only,
respectively.
}
\label{tab:cov_modes}
\begin{tabular}{lccccc}
\toprule
Comparison
& $N_{\rm kept}$
& $\lambda_2/\lambda_1$
& Leading-mode fraction
& $S_{\rm hist}^{(2)}$
& $S_{\rm hist}^{(1)}$ \\
\midrule
DESI DR2 vs Planck
& $2$
& $3.3\times10^{-6}$
& $0.989$
& $1.65$
& $1.64$ \\
Pantheon$+$SH0ES vs Planck
& $2$
& $6.7\times10^{-6}$
& $0.792$
& $2.55$
& $2.27$ \\
\bottomrule
\end{tabular}
\end{table}

For DESI DR2 vs Planck, the leading retained mode accounts for about
$98.9\%$ of $S_{\rm hist}^2$, and the one-mode and two-mode displacements are
nearly identical. For Pantheon$+$SH0ES vs Planck, the leading mode accounts for
about $79.2\%$ of $S_{\rm hist}^2$, so the second retained mode contributes
non-negligibly. In both cases, the qualitative conclusion is unchanged if the
comparison is restricted to the dominant mode.

The eigenmode decomposition clarifies the character of the mismatch. The
history-space difference is not a collection of independent redshift-bin
tensions. It is a coherent displacement in a low-dimensional covariance
subspace. The reconstructed dimensionless histories therefore differ across
probes, but the discrepancy remains mild-to-moderate once the redshift
covariance is included. In the normalized history variable $E(z)$, the
inter-probe mismatch is smaller than the conventional scalar-$H_0$ comparison.

%==========================================================
\section{Implications for the Hubble Tension}
\label{sec:4}
%==========================================================

The results above separate two statistical questions. The first is the usual one-dimensional comparison of quoted scalar values of $H_0$. The second is the comparison of posterior-implied normalized expansion histories, $E(z)$. These comparisons need not give the same level of mismatch.

In the flat-$\Lambda$CDM projection used here, the mismatch in $E(z)$ is
mild-to-moderate. The fractional differences are at the percent level, and the
pointwise significances are typically of order unity. After the redshift
covariance is included, the fiducial two-mode covariance-subspace displacements
are $S_{\rm hist}^{(2)}\simeq 1.65$ for DESI DR2 vs Planck and
$S_{\rm hist}^{(2)}\simeq 2.55$ for Pantheon$+$SH0ES vs Planck. For
$N_{\rm kept}=2$, these correspond to two-sided one-dimensional
Gaussian-equivalent significances of approximately $1.1\sigma$ and
$2.1\sigma$, respectively. These Gaussian-equivalent values are smaller than
the conventional $\sim4.9\sigma$ scalar Planck--SH0ES discrepancy
~\cite{Planck:2018vyg,Verde:2019ivm,Riess:2021jrx}.

This result does not remove the Hubble tension. It also does not remove the absolute-scale discrepancy encoded in the distance ladder. It shows a more specific point: the scalar-$H_0$ discrepancy and the normalized-history discrepancy are different diagnostics. A large mismatch in the quoted value of $H_0$ does not automatically imply an equally large mismatch in the reconstructed dimensionless expansion history.

The use of $E(z)$ removes the explicit absolute normalization, but it does not make the comparison trivial. In flat $\Lambda$CDM, the remaining normalized history is controlled mainly by $\Omega_{m0}$, whose relation to $H_0$ is probe dependent. Planck constrains physical-density combinations such as $\Omega_{m0}h^2$ and the acoustic scale, while the distance ladder fixes the absolute scale through calibration. Thus, agreement or disagreement in $E(z)$ is a statement about the normalized expansion shape implied by different inference chains, not an algebraic consequence of dividing by $H_0$.

The comparison in $E(z)$ should be viewed as a scale--shape decomposition. The scalar $H_0$ comparison diagnoses the absolute normalization of the late-time distance scale. The $E(z)$ comparison diagnoses the redshift-dependent shape of the expansion history. A large discrepancy in $E(z)$ would indicate a failure of the background expansion shape itself. The present result shows instead that, within this flat-$\Lambda$CDM projection test, the shape-level mismatch is much weaker than the absolute-scale discrepancy.

The distinction matters because the three probes do not constrain the same primitive observable. Planck CMB constrains acoustic information and physical densities.  DESI DR2 BAO constrains ruler-normalized distances. Pantheon$+$SH0ES constrains a calibrated luminosity-distance relation. The quoted scalar $H_0$ is obtained only after these quantities are projected through a background model and, in the distance-ladder case, through an absolute calibration~\cite{Knox:2019rjx,Efstathiou:2021ocp,Brieden:2022heh,Verde:2023lmm}.  The same symbol $H_0$ therefore denotes derived quantities obtained through different inference chains. This projection dependence is closely related to the information-geometric viewpoint, where the model manifold controls the orientation and sharpness of the inferred late-time constraints~\cite{Lee:2026ets}.

The present analysis keeps the model class fixed. All histories are constructed within flat base $\Lambda$CDM. This is not a model-independent reconstruction. It is a controlled common-model comparison. The purpose is to ask how the three probe posteriors map into the same dimensionless history space when the background model is held fixed.

This choice has a consequence. In flat $\Lambda$CDM, the shape of $E(z)$ is controlled mainly by $\Omega_{m0}$. Therefore, the reconstructed values $E(z_k)$ on the redshift grid are strongly correlated. A diagonal redshift-by-redshift comparison over-counts the same coherent displacement. The covariance-aware statistic instead measures the displacement in the posterior-supported covariance subspace.

The rigidity can be seen analytically. Since $E(z)$ is given by Eq.~\eqref{Ez},
\begin{equation}
\frac{\partial E(z)}{\partial \Omega_{m0}} = \frac{(1+z)^3-1}{2E(z)} ,
\label{eq:dEdOm}
\end{equation}
and, to leading order,
\begin{equation}
\sigma_E^2(z) \simeq \left[ \frac{(1+z)^3-1}{2E(z)} \right]^2 \sigma^2_{\Omega_{m0}} . \label{eq:sigmaE_prop}
\end{equation}
Thus, the apparent precision of the reconstructed $E(z)$ curve is not produced by independent measurements of $H(z)$ at many redshifts. It is the propagated precision of a low-dimensional background model. This is the sense in which flat $\Lambda$CDM imposes a rigid expansion-history manifold.

The result should be interpreted with this limitation in mind. It does not say that the absolute value of $H_0$ is unimportant. It says that the nearly five-sigma scalar Planck--SH0ES discrepancy is not mirrored by a comparably large discrepancy in normalized expansion history. The two statements concern different projections of the same inference problem.

Table~\ref{tab:Ez_tension_summary} summarizes the pointwise mismatch measures and the fiducial two-mode covariance-subspace history displacements. The pointwise significances describe the largest local offsets on the chosen redshift grid. By contrast, $S_{\rm hist}^{(2)}$ is a global displacement in the retained covariance eigenmode subspace. It should not be read as a one-dimensional number of sigma. The corresponding one-dimensional Gaussian equivalents are approximately $1.1\sigma$ for DESI DR2 vs Planck and $2.1\sigma$ for Pantheon$+$SH0ES vs Planck. As discussed in Section~\ref{sec:3}, the covariance spectrum is highly hierarchical, so the two-mode result should be understood as a regularized covariance-subspace choice, not as evidence for two independent redshift-bin measurements.

\begin{table*}[t]
\centering
\caption{Summary of pointwise mismatch measures and fiducial two-mode covariance-subspace history displacements for the reconstructed $E(z)$ histories relative to Planck. The final column is not a frequentist significance from independent redshift bins; it is the displacement in the retained covariance eigenmode subspace. The leading-mode-only stability check is given in Table~\ref{tab:cov_modes}.}
\label{tab:Ez_tension_summary}
\begin{tabular}{lcccc}
\toprule
Comparison
& Max frac.\ mismatch
& Max pointwise significance
& $N_{\rm kept}$
& $S_{\rm hist}^{(2)}$ \\
\midrule
DESI DR2 vs Planck
& $\sim 3.1\%$
& $\sim 1.68\sigma$
& $2$
& $\sim 1.65$ \\
Pantheon$+$SH0ES vs Planck
& $\sim 6.4\%$
& $\sim 2.29\sigma$
& $2$
& $\sim 2.55$ \\
\bottomrule
\end{tabular}
\end{table*}

Thus, within the flat-$\Lambda$CDM projection used here, the probe posteriors show a weaker mismatch in the normalized shape-level history than in the absolute-scale scalar-$H_0$ projection. The result does not explain the origin of the distance-ladder discrepancy; it identifies how the discrepancy is distributed between scale and shape.

%==========================================================
\section{Discussion and Conclusion}
\label{sec:5}
%==========================================================

This paper has examined the Hubble tension at the level of reconstructed dimensionless background history. The starting point is that Planck CMB, DESI DR2 BAO, and Pantheon$+$SH0ES do not constrain the same primitive observable. Each probe constrains different probe-level quantities. The quoted scalar $H_0$ is therefore a derived late-time summary, not a common primary observable.

We reconstructed posterior-implied $E(z)$ histories for the three probes within the same flat base $\Lambda$CDM framework. The histories do not coincide exactly. DESI DR2 and Pantheon$+$SH0ES show visible offsets relative to Planck. However, the mismatch is smaller than the conventional scalar-$H_0$ comparison. With the redshift covariance included, the fiducial two-mode covariance-subspace displacements are
\[
S_{\rm hist}^{(2)}\simeq 1.65
\quad \text{for DESI DR2 vs Planck},
\qquad
S_{\rm hist}^{(2)}\simeq 2.55
\quad \text{for Pantheon$+$SH0ES vs Planck}.
\]
For $N_{\rm kept}=2$, these correspond to two-sided one-dimensional Gaussian-equivalent significances of approximately $1.1\sigma$ and $2.1\sigma$, respectively.

The covariance treatment is essential. In flat $\Lambda$CDM, the reconstructed history is highly correlated across redshift. A diagonal sum over redshift bins therefore over-counts the same coherent displacement. The covariance-aware comparison instead measures the displacement of the reconstructed history in the retained covariance eigenmode subspace. Since this subspace is nearly rank-one, $N_{\rm kept}=2$ should be interpreted as a regularized covariance-subspace choice, not as evidence for two independent redshift-bin measurements.

The conclusion is limited. The Hubble tension does not disappear, and the absolute-scale discrepancy remains. Within the flat-$\Lambda$CDM projection used here, however, the same probe posteriors show only a mild-to-moderate mismatch in the normalized history variable $E(z)$. Thus the scalar-$H_0$ comparison is sharper than the corresponding mismatch in reconstructed dimensionless background history.

This distinction matters for interpreting cosmological tensions. A scalar parameter discrepancy is not automatically a history-level discrepancy. It may reflect a difference in the reconstructed background history, or it may reflect the way different probe-level observables are projected into derived late-time parameters. The present analysis separates these two possibilities within the simplest common framework by decomposing the expansion comparison into an absolute-scale component and a normalized-history, or shape, component.

The present analysis deliberately keeps the model space fixed to flat $\Lambda$CDM. This choice isolates the distinction between the scalar-$H_0$ projection and the normalized-history projection in the model where the standard Hubble tension is usually quoted. Extending the comparison to $w$CDM or $w_0w_a$CDM would address a different problem, rather than a simple completion of the present one. In those models, additional late-time degrees of freedom can broaden and rotate the projected posterior in $H_0$, and distance data alone have limited power to constrain the time evolution of the dark energy sector. This can dilute the scalar-$H_0$ discrepancy, but it also changes the model manifold being tested. For this reason, we do not interpret the extended-model case as a direct test of the flat-$\Lambda$CDM result presented here. It is instead a separate question: how the scale--shape decomposition changes when the background manifold itself is enlarged~\cite{Lee:2026ets,Lee:2025jrr,Lee:2026sta}.

%============================================================
\appendix
\section{Numerical Methodology for the Reconstructed $E(z)$ Histories}
\label{app:numerical_method}
%============================================================

This appendix describes the numerical procedure used to construct the
dimensionless expansion histories \(E(z)\) and the corresponding mismatch
diagnostics. The analysis uses three inputs: Planck 2018 base-\(\Lambda\)CDM
posterior chains, DESI DR2 BAO constraints, and the Pantheon$+$SH0ES supernova
data set. Table~\ref{tab:numerical_method} summarizes the numerical setup.

All reconstructions are performed within flat base \(\Lambda\)CDM. The working
expression is Eq.~\eqref{Ez}. The radiation density is not sampled independently
in the numerical implementation. It is omitted in the late-time reconstruction.
This approximation is sufficient for the redshift range used in the main
comparison.

%------------------------------------------------------------
\subsection{Planck 2018}
%------------------------------------------------------------

For Planck, we use the public Planck 2018 base-\(\Lambda\)CDM posterior chains.
These chains provide weighted posterior samples of cosmological and derived
parameters. They are not raw CMB power-spectrum data.

We extract \(\Omega_{m0}\) and the corresponding sample weights. A burn-in
fraction of \(30\%\) is removed before constructing the reconstructed histories.
For each retained sample, Eq.~\eqref{Ez} is evaluated on the common redshift
grid. The median history, credible bands, and redshift covariance are estimated
from the weighted sample ensemble.

%------------------------------------------------------------
\subsection{DESI DR2 BAO}
%------------------------------------------------------------

For DESI DR2, we use the BAO observables and their covariance information. The
relevant observables include \(D_M/r_d\) and \(D_H/r_d\), with isotropic
combinations used where applicable. In the present implementation, the DESI
posterior is represented by a Gaussian approximation in the flat-\(\Lambda\)CDM
parameter space. The approximation is defined by the best-fit vector and Fisher
matrix used in this analysis.

Posterior samples are drawn from the corresponding multivariate normal
distribution. Each sample is mapped to \(E(z)\) through its sampled value of
\(\Omega_{m0}\) using Eq.~\eqref{Ez}. The DESI redshift covariance of \(E(z)\)
is computed from the resulting sample ensemble.

This procedure is used to place DESI DR2 on the same flat-\(\Lambda\)CDM
comparison footing as the other probes. It is not a substitute for a full DESI
likelihood analysis.

The Gaussian approximation is used here only to propagate the local posterior
width and covariance of the flat-\(\Lambda\)CDM parameters into the reconstructed
\(E(z)\) ensemble. The final statistic is not a tail probability computed from
the DESI likelihood itself. Possible non-Gaussian tails of the full BAO
posterior are therefore sub-leading for the present purpose, which is to compare
the central posterior-implied histories and their covariance-supported
directions in a common model space.

%------------------------------------------------------------
\subsection{Pantheon$+$SH0ES}
%------------------------------------------------------------

For Pantheon$+$SH0ES, we use the supernova data vector and covariance matrix
together with the SH0ES calibration. The theoretical luminosity distance is
given by Eq.~\eqref{DL}. The theoretical distance modulus is written as
\begin{equation}
\mu_{\rm th}(z)
=
5\log_{10}\!\left[\frac{D_L(z)}{\rm Mpc}\right]
+25+M,
\label{eq:mu_app}
\end{equation}
where \(M\) denotes the calibration parameter used in the implementation. The
function \(E(z')\) entering Eq.~\eqref{DL} is evaluated using Eq.~\eqref{Ez}.

The best-fit point is obtained in the parameter space
\((H_0,\Omega_{m0},M)\). We then construct a Gaussian approximation to the
posterior around that point. The derivatives of the theoretical distance
modulus are evaluated by central finite differences. With the Jacobian matrix
\(\mathbf{J}\), the approximate Fisher matrix is
\begin{equation}
\mathbf{F}_{\rm SN}
=
\mathbf{J}^{\mathsf{T}}
\mathbf{C}_{\rm SN}^{-1}
\mathbf{J}
+
\mathbf{F}_{\rm prior}.
\label{eq:Fisher_SN_app}
\end{equation}
Here \(\mathbf{C}_{\rm SN}\) is the Pantheon$+$SH0ES covariance matrix, and
\(\mathbf{F}_{\rm prior}\) represents the Gaussian prior on the calibration
parameter. Explicitly, for the parameter vector
\[
{\boldsymbol\theta}=(H_0,\Omega_{m0},M),
\]
we use
\begin{equation}
\mathbf{F}_{\rm prior}
=
\mathrm{diag}
\left(
0,\,
0,\,
\sigma_M^{-2}
\right),
\label{eq:Fprior_M}
\end{equation}
where \(\sigma_M\) is the adopted uncertainty of the SH0ES absolute-magnitude
calibration. In the numerical results reported here we use
\begin{equation}
\sigma_M = 0.03~{\rm mag}.
\label{eq:sigmaM_prior}
\end{equation}

Posterior samples are drawn from the Gaussian approximation. Each sample is
mapped to \(E(z)\) through \(\Omega_{m0}\) using Eq.~\eqref{Ez}. The redshift
covariance of the reconstructed Pantheon$+$SH0ES history is computed from this
sample ensemble.

This Gaussian approximation is likewise used only as a local covariance
propagation scheme. Since the reconstructed \(E(z)\) history in flat
\(\Lambda\)CDM depends only on \(\Omega_{m0}\), the relevant requirement is that
the posterior width of \(\Omega_{m0}\) be represented accurately near the
posterior maximum. The covariance-subspace displacement \(S_{\rm hist}\) is
therefore insensitive to far-tail non-Gaussianities at leading order. A full
non-Gaussian posterior sampling of Pantheon$+$SH0ES would be required for a
precision tail-probability calculation, but that is not the role of
\(S_{\rm hist}\) in this work.

%------------------------------------------------------------
\subsection{Pointwise Measures and Covariance-Subspace Displacement}
%------------------------------------------------------------

The pointwise significance shown in Fig.~\ref{fig:Ez_reconstruction} is given
by Eq.~\eqref{eq:pointwise_significance}. The quantity \(\sigma(z)\) is
estimated from the \(68\%\) posterior interval of each reconstructed history.
This statistic is used for the pointwise panel of
Fig.~\ref{fig:Ez_reconstruction}.

A diagonal redshift-by-redshift statistic is given by Eq.~\eqref{eq:chi2_diag}.
This expression over-counts the mismatch when the reconstructed histories are
strongly correlated across redshift. Therefore, it is not used as a physical
tension measure. It is used only as a diagnostic of the over-counting produced
by treating correlated redshift samples as independent.

The covariance-aware comparison uses the redshift covariance induced by the
posterior samples. For each probe, the sample histories define vectors
\begin{equation}
{\bf E}^{(a)}
=
\left(
E^{(a)}(z_1),
E^{(a)}(z_2),
\ldots,
E^{(a)}(z_N)
\right),
\label{eq:E_vector_app}
\end{equation}
where \(a\) labels a posterior sample. The mean vector is
\begin{equation}
\bar{\bf E}
=
\frac{1}{N_s}
\sum_{a=1}^{N_s}
{\bf E}^{(a)} .
\label{eq:E_mean_app}
\end{equation}
The covariance matrix is
\begin{equation}
C_E
=
\frac{1}{N_s-1}
\sum_{a=1}^{N_s}
\left({\bf E}^{(a)}-\bar{\bf E}\right)
\left({\bf E}^{(a)}-\bar{\bf E}\right)^{\mathsf T}.
\label{eq:E_cov_app}
\end{equation}

For two independent probes, the covariance of the difference is given by
Eq.~\eqref{eq:Cdelta_main}. For the difference vector
\[
\Delta{\bf E}
=
\bar{\bf E}_i-\bar{\bf E}_{\rm Planck},
\]
the covariance-aware statistic is
\begin{equation}
\chi_E^2
=
\Delta{\bf E}^{\mathsf T}
C_{\Delta E}^{+}
\Delta{\bf E}.
\label{eq:chi2_cov_app}
\end{equation}
In practice, we diagonalize \(C_{\Delta E}\) as
\begin{equation}
C_{\Delta E}
=
\mathbf{U}
\,\mathrm{diag}(\lambda_1,\lambda_2,\ldots,\lambda_N)
\,\mathbf{U}^{\mathsf T},
\label{eq:cov_eigendecomp_app}
\end{equation}
with eigenvalues ordered from largest to smallest. The displacement vector is
then projected onto the covariance eigenbasis,
\begin{equation}
\Delta{\bf e}
=
\mathbf{U}^{\mathsf T}\Delta{\bf E}.
\label{eq:Delta_e_app}
\end{equation}

Because the covariance spectrum is highly hierarchical, the retained-rank choice should be regarded as a regularization of the covariance subspace, not as a count of independent redshift-bin measurements. Only eigenmodes with non-negligible posterior-supported variance are retained. The inverse covariance used in Eq.~\eqref{eq:chi2_cov_app} is therefore the pseudo-inverse on the retained subspace,
\begin{equation}
\chi_E^2
=
\sum_{\alpha\in{\rm kept}}
\frac{(\Delta e_\alpha)^2}{\lambda_\alpha}.
\label{eq:chi2_eigen_app}
\end{equation}
We report the covariance-subspace history displacement
\begin{equation}
S_{\rm hist}
=
\sqrt{\chi_E^2}.
\label{eq:Shist_app}
\end{equation}
This quantity is an effective displacement in the retained covariance
eigenmode subspace. It is not a frequentist significance obtained by treating
the redshift grid points as independent data.

The fractional contribution of each retained eigenmode to the total
covariance-subspace displacement is computed at the level of
\(S_{\rm hist}^2=\chi_E^2\):
\begin{equation}
f_\alpha
=
\frac{(\Delta e_\alpha)^2/\lambda_\alpha}
{\sum_{\beta\in{\rm kept}}(\Delta e_\beta)^2/\lambda_\beta}.
\label{eq:eigenmode_fraction_app}
\end{equation}
The leading-mode fraction quoted in Table~\ref{tab:cov_modes} is the largest
of these retained-mode contributions.

%------------------------------------------------------------
\subsection{Gaussian-Equivalent Interpretation of $S_{\rm hist}$}
\label{app:gaussian_equivalent}
%------------------------------------------------------------

This subsection clarifies how the covariance-subspace displacement $S_{\rm hist}$ is related to a $p$-value and to a one-dimensional Gaussian-equivalent significance. This conversion is used only as an interpretive aid. The primary quantity reported in this work is $S_{\rm hist}$ itself.

For a one-dimensional standard normal variable $Z$, the probability density is
\begin{equation}
\phi(z) = \frac{1}{\sqrt{2\pi}} \exp\left(-\frac{z^2}{2}\right), \label{eq:normal_pdf_app}
\end{equation}
where $Z$ denotes the random variable and $z$ denotes a possible numerical value of that variable. The cumulative distribution function is
\begin{equation}
\Phi(z) = \int_{-\infty}^{z}\phi(t)\,dt . \label{eq:normal_cdf_app}
\end{equation}
The value of the density $\phi(z)$ at a point is not a tail probability. A two-sided one-dimensional Gaussian significance $Z_{\rm 1D}$ is defined by
\begin{equation}
p = 2\left[1-\Phi(Z_{\rm 1D})\right]. \label{eq:p_to_z1d_app}
\end{equation}

The covariance-subspace statistic used in this paper is different. After diagonalizing the covariance matrix, we retain only the eigenmodes with non-negligible posterior-supported variance. The set of these modes is denoted by \rm{kept}. This is the same logic as a covariance eigenmode, or Karhunen--Loeve, compression: the data vector is rotated into the eigenbasis of its covariance, and each retained component is rescaled by its standard deviation~\cite{Tegmark:1997rp,Heavens:2009nx}. The resulting coordinates are called whitened coordinates because, in the retained subspace, their covariance is the identity matrix.

For each retained mode, we define
\begin{equation}
y_\alpha = \frac{\Delta e_\alpha}{\sqrt{\lambda_\alpha}}, \qquad \alpha\in{\rm{kept}} \,,
\label{eq:whitened_modes_app}
\end{equation}
where $\Delta e_\alpha$ is the component of the history difference along the $\alpha$th covariance eigenmode, and $\lambda_\alpha$ is the variance along that mode. Thus, $y_\alpha$ is the standardized pull in that eigenmode.

Using Eq.~\eqref{eq:chi2_eigen_app}, one has
\begin{equation}
S_{\rm hist}^2 \equiv \chi_E^2 = \sum_{\alpha\in{\rm kept}} y_\alpha^2 . \label{eq:Shist_chi2_app}
\end{equation}
If the retained modes are treated as Gaussian degrees of freedom, then $S_{\rm hist}^2$ follows a chi-square distribution with
\begin{equation}
\nu=N_{\rm kept} \label{eq:nu_Nkept_app}
\end{equation}
degrees of freedom under the null hypothesis of consistent histories. The corresponding tail probability is
\begin{equation}
p_{\rm hist} = P\!\left(\chi_\nu^2\ge S_{\rm hist}^2\right) = 1-F_{\chi_\nu^2}\!\left(S_{\rm hist}^2\right), \label{eq:phist_app}
\end{equation}
where $F_{\chi_\nu^2}$ is the cumulative distribution function of the chi-square distribution. Equivalently,
\begin{equation}
p_{\rm hist} = \frac{\Gamma\!\left(\nu/2,S_{\rm hist}^2/2\right)}{\Gamma\!\left(\nu/2\right)} . \label{eq:phist_gamma_app}
\end{equation}
For the case used in the main analysis, $N_{\rm kept}=2$, this reduces to
\begin{equation}
p_{\rm hist} = \exp\left(-\frac{S_{\rm hist}^2}{2}\right). \label{eq:phist_df2_app}
\end{equation}

The one-dimensional Gaussian-equivalent significance is then obtained by solving
\begin{equation}
p_{\rm hist} = 2\left[1-\Phi(Z_{\rm 1D})\right], \label{eq:Z1D_from_phist_app}
\end{equation}
or
\begin{equation}
Z_{\rm 1D} = \Phi^{-1}\!\left(1-\frac{p_{\rm hist}}{2}\right). \label{eq:Z1D_inverse_app}
\end{equation}
Thus, for $N_{\rm kept}=2$,
\begin{align}
S_{\rm hist}=1.65 &\quad\Rightarrow\quad \chi_E^2\simeq2.72, \quad p_{\rm hist}\simeq0.26, \quad
Z_{\rm 1D}\simeq1.1\sigma, \label{eq:Shist_165_app} \\
S_{\rm hist}=2.55 &\quad\Rightarrow\quad \chi_E^2\simeq6.50, \quad p_{\rm hist}\simeq0.039, \quad
Z_{\rm 1D}\simeq2.1\sigma. \label{eq:Shist_255_app}
\end{align}

It is also useful to distinguish the one-dimensional Gaussian equivalent from a two-dimensional enclosed-probability contour. For $\nu=2$, the probability inside a radius $S$ in the retained covariance subspace is
\begin{equation}
P(<S) = 1-\exp\left(-\frac{S^2}{2}\right). \label{eq:twoD_enclosed_app}
\end{equation}
Therefore, the radius enclosing a probability $P$ is
\begin{equation}
S(P) = \sqrt{-2\ln(1-P)}. \label{eq:twoD_radius_app}
\end{equation}
The usual one-, two-, and three-sigma enclosed probabilities of a one-dimensional Gaussian, $P=0.6827$, $0.9545$, and $0.9973$, correspond in two dimensions to
\begin{equation}
S\simeq1.52,\qquad 2.49,\qquad 3.44, \label{eq:twoD_sigma_radii_app}
\end{equation}
respectively. Hence a radial displacement $S_{\rm hist}$ should not be read as a one-dimensional number of sigma. The conversion must pass through $p_{\rm hist}$, as in Eqs.~\eqref{eq:phist_app}--\eqref{eq:Z1D_inverse_app}.

%------------------------------------------------------------
\subsection{Redshift Grid and Numerical Settings}
%------------------------------------------------------------

All reconstructed histories are evaluated on the same redshift grid. This means that each posterior-implied expansion history is sampled at the same set of redshift values, $z_1,z_2,\cdots,z_N$. Using a common grid ensures that the histories from Planck, DESI DR2, and Pantheon$+$SH0ES can be compared component by component. The same grid is used to compute the median histories, credible bands, pointwise significance, and covariance-aware comparison. The median history is the pointwise median of the posterior-implied $E(z)$ ensemble. The credible band is the corresponding posterior interval at each redshift. The pointwise significance compares the local offset between two mean histories at a single redshift, whereas the covariance-aware comparison uses the full redshift covariance and therefore accounts for the strong correlations among the reconstructed values $E(z_k)$.

For Planck, a burn-in fraction of $30\%$ is removed before the reconstructed histories are computed. This means that the first $30\%$ of the Markov-chain samples are discarded to reduce the influence of the initial sampling transient. The remaining weighted samples are then used to construct the posterior-implied $E(z)$ ensemble.

For DESI DR2, the posterior is represented by a Gaussian approximation in the flat-\(\Lambda\)CDM parameter space. This approximation is specified by the best-fit parameter vector and Fisher matrix used in the analysis. Posterior samples are drawn from the corresponding multivariate normal distribution and then mapped to \(E(z)\) through the sampled value of \(\Omega_{m0}\) using Eq.~\eqref{Ez}. In contrast to the Pantheon$+$SH0ES Fisher construction below, no finite-difference derivatives of a distance-modulus model are used at this stage. The resulting ensemble of DESI-implied \(E(z)\) curves is used to compute the median history, credible band, and redshift covariance.

For the Pantheon$+$SH0ES Gaussian approximation, derivatives of the theoretical distance modulus with respect to the parameters $(H_0,\Omega_{m0},M)$ are computed by central finite differences. The finite-difference step sizes are
\begin{equation}
\Delta H_0 = 10^{-3}, \qquad \Delta \Omega_{m0}=10^{-4}, \qquad \Delta M = 10^{-4} \,, \label{eq:steps_app}
\end{equation}
where $M$ denotes the supernova absolute-magnitude calibration parameter used in the likelihood implementation. The step sizes are numerical increments used only to estimate derivatives; they are not physical priors or observational uncertainties.

\begin{table*}[t]
\centering
\caption{Numerical setup used to reconstruct the \(E(z)\) histories. Planck is reconstructed from weighted posterior chains. DESI DR2 and Pantheon$+$SH0ES are represented by Gaussian posterior approximations within flat \(\Lambda\)CDM. The late-time histories are evaluated with Eq.~\eqref{Ez}.}
\label{tab:numerical_method}

\footnotesize
\setlength{\tabcolsep}{4pt}
\renewcommand{\arraystretch}{1.15}

\begin{tabularx}{\textwidth}{p{3.2cm}
>{\RaggedRight\arraybackslash}X
>{\RaggedRight\arraybackslash}X
>{\RaggedRight\arraybackslash}X}
\hline\hline
 & Planck 2018 & DESI DR2 BAO & Pantheon$+$SH0ES \\
\hline

Input data
& Public base-\(\Lambda\)CDM posterior chains
& BAO observables and covariance
& Pantheon$+$SH0ES data vector and covariance \\

Reconstruction basis
& Weighted posterior samples in flat \(\Lambda\)CDM
& Gaussian posterior approximation in flat \(\Lambda\)CDM
& Gaussian posterior approximation in \((H_0,\Omega_{m0},M)\) \\

Posterior generation
& Directly from weighted chains
& Multivariate normal sampling from best fit and Fisher matrix
& Multivariate normal sampling from best fit and Fisher matrix \\

Finite differences
& Not used
& Not used
& Central finite differences in \((H_0,\Omega_{m0},M)\) \\

Step sizes
& N/A
& N/A
& \(\Delta H_0 = 10^{-3}\),
\(\Delta \Omega_{m0} = 10^{-4}\),
\(\Delta M = 10^{-4}\) \\

Burn-in
& \(30\%\)
& Not applicable
& Not applicable \\

History covariance
& Posterior sample covariance of \(E(z)\)
& Posterior sample covariance of \(E(z)\)
& Posterior sample covariance of \(E(z)\) \\

Covariance-subspace statistic
& \multicolumn{3}{>{\centering\arraybackslash}p{0.72\textwidth}}%
{\(\chi_E^2=\Delta{\bf E}^{\mathsf T}C_{\Delta E}^{+}\Delta{\bf E}\),
evaluated with the pseudo-inverse on the retained covariance-eigenmode
subspace; \(S_{\rm hist}=\sqrt{\chi_E^2}\)} \\

\hline\hline
\end{tabularx}
\end{table*}

%================================================================================

\end{document}